\documentclass{article}
\usepackage[utf8]{inputenc}
\usepackage{graphicx}
\usepackage{hyperref}
\usepackage{geometry}
\usepackage{booktabs}
\title{Evolving the Productivity Equation: Should Digital Labor Be Considered a New Factor of Production?}
\author{Alex Farach, Alexia Cambon, Jared Spataro}
\date{May 13, 2025}

\begin{document}

\maketitle

\begin{abstract}
As the digital economy grows increasingly intangible, traditional productivity measures struggle to capture the true economic impact of artificial intelligence (AI). AI systems capable of cognitive work significantly enhance productivity, yet their contributions remain obscured within the residual category of Total Factor Productivity (TFP). This paper explores whether it is time for a conceptual shift to explicitly recognize "digital labor,'' the autonomous cognitive capability of AI, as a distinct factor of production in addition to traditional capital and human labor. We outline the unique economic properties of digital labor, such as unprecedented scalability, intangibility, self-improvement through autonomous learning, rapid obsolescence, and nuanced substitutability with human labor. By explicitly integrating digital labor into economic growth models (for example, the Solow and Romer frameworks), we demonstrate critical strategic implications for business leaders, including new approaches to productivity tracking, resource allocation, investment strategy, and organizational design. Ultimately, distinguishing digital labor as an independent factor can provide clearer insight into economic growth and position organizations to effectively manage and leverage AI’s transformative potential.
\end{abstract}

\vspace{0.5em}

\section{Introduction}
Since the commercial internet took hold in the mid‑1990s, production has been steadily dematerializing. Music, maps, advertising, retail catalogs, and even portions of industrial design now reach customers as a weightless code. Intermediaries disappear; platforms deliver free services that are paid for, if at all, through data flows and attention markets. Software running inside datacenter racks substitutes for hardware once visible on factory floors. In a world where goods, services, and value chains are increasingly intangible and globally distributed, counting widgets produced is an ever-poorer proxy for counting progress. The statistical machinery that converts nominal spending into 'real' output: price indices, quality deflators, hedonic adjustments, struggles to keep pace, leaving policymakers navigating an economic fog \cite{coyle2025measure}.

Artificial intelligence (AI) will densify that fog. Systems capable of cognitive work at scale will spread productivity gains across organizations, yet leave few physical traces. If we continue to treat AI as an unseen factor folded into Total Factor Productivity (TFP), we risk widening the gap between what we measure and what we need to know. A large body of research already shows that AI systems boost output, quality, and speed across occupations \cite{jaffe2024microsoft}; yet the newest foundation models and agentic capabilities extend those gains to a broader set of economically viable cognitive tasks. Their scale, versatility, and rapid diffusion suggest that we have reached a point where intelligence deserves separate treatment in our production accounting. 

This paper explores whether digital labor, the autonomous cognitive capability of AI, could be considered as a distinct factor of production, along with capital and human labor. We will explore three key themes:

\begin{enumerate}
    \item The Productivity Measurement Problem: Why current productivity metrics and accounting methods are insufficient to capture the economic impact of AI and other intangible, knowledge-based value creation. We will see how relying on TFP as a catch-all obscures what is really happening and why our national accounts and KPIs are struggling to value AI-driven output.
    \item Digital Labor as a New Factor: Why could AI - or digital labor - be recognized as an independent input to production, distinct from traditional capital and human labor? We will highlight its unique economic characteristics, scalability, intangibility, self-improvement, rapid depreciation, and elastic roles, and show why neither the 'capital' nor 'human labor' categories fully capture its behavior.
    \item Strategic Implications for Business: How explicitly integrating digital labor into growth models, including Solow’s model of capital accumulation and Romer’s model of innovation-driven growth, changes the strategy for business leaders. We will translate these theoretical insights into practical guidance on resource allocation, productivity tracking, investment strategy, and organizational design.

\end{enumerate}

By understanding digital labor as a production factor, business leaders can better measure its contributions and more effectively manage its integration into their businesses.

\section{Why Traditional Productivity Metrics Fall Short in the AI Age}
Despite rapid technological advances, official productivity statistics in many industries and economies remain sluggish. Part of the reason is that our metrics were not designed for the digital, intangible economy. Gross domestic product (GDP) and classical productivity measures were developed in an era when physical goods and manual labor dominated, where output was easily counted and value was tied to tangible products. Today, however, a growing share of economic value comes from intangible outputs – software, data, models, designs, customer experiences – which are harder to quantify \cite{coyle2025measure}. AI further complicates this picture by generating value in ways that traditional accounting and statistics often miss.

\subsection{Intangible outputs and “free” services}
Take a generative AI system that reduces delivery times by optimizing logistics, or one that provides real-time customer support. Even when these services lead to substantial improvements in speed or quality, their economic contribution may remain undercounted, especially if the price charged does not reflect the magnitude of improvement. That’s because traditional GDP only counts what is transacted, not what is transformed \cite{brynjolfsson2017paradox}. Similarly, AI that improves work quality or productivity, without altering headcount or capital stock, often disappears into the shadows of firm-level KPIs.

Even within organizations, the value of AI tools may be economically material, but statistically invisible. Accounting standards such as IAS 38 exclude most internally developed AI systems from balance sheets unless they meet strict capitalization criteria \cite{iasb2021ias38}. As a result, highly productive models can be treated as expenses rather than investments, underestimating their long-term value. This dynamic is part of what Haskel and Westlake call the 'intangible economy', where assets such as software, data, and AI models generate returns without appearing in the ledgers \cite{haskel2017capitalism}.

\subsection{Total Factor Productivity (TFP) as a black box}
Economists traditionally lump the effects of new technology into a residual called TFP - a residual term that captures everything not explained by capital or labor inputs \cite{solow1957technical}. TFP is sometimes dubbed a measure of 'technological progress', but treating the impact of AI this way hides critical information. TFP is fundamentally a residual, a 'black box' offering no insight into how productivity improvements occur. Was it due to better worker skills? More efficient machines? Or an AI model optimizing operations? If AI-driven gains are buried in TFP, business leaders and policymakers can’t tell. Relying on aggregate TFP to gauge AI’s impact is like measuring a car’s performance with a single gauge that mysteriously increases, without knowing if the engine, the fuel, or the tires made the difference.

\subsection{National accounting lags}
The frameworks supporting GDP and productivity statistics are outdated for a digital world \cite{coyle2025measure}  \cite{haskel2017capitalism} \cite{brynjolfsson2017paradox}. The “economic fog” is thickening – our statistics struggle to convert nominal spending into “real” output when quality is improving rapidly or when output is a digital good. Government agencies find it difficult to measure the productivity of AI that writes code or diagnoses diseases because these contributions do not fit neatly into existing industry categories or price indexes. Similarly, within firms, KPIs may not capture AI contributions \cite{humlum2025llms}. For example, if an AI assistant enables a team to handle 30\% more customer inquiries in a day, the boost might be attributed vaguely to “efficiency” without clearly showing AI’s role.

In short, current productivity metrics undercount AI-driven gains and intangible value creation. They were built for an economy of physical things, but today’s economy runs on digital information. A new approach to measurement is needed, one that can identify and credit the contributions of AI. That is a key motivation for pulling digital labor out of the TFP shadows and considering it as its own factor of production. We can directly track how much AI (as digital labor) we are investing in and what output it generates, rather than letting it hide in a residual. 

\section{Digital Labor as a Third Factor of Production Distinct from Capital and Human Labor}
In this context, digital labor refers to the autonomous capabilities of AI systems to perform cognitive tasks that drive economic output, such as handling customer inquiries, optimizing supply chains, designing components or accelerating drug discovery. But where do these capabilities fit within the traditional production framework? Are they best understood as capital or do they represent something fundamentally new?

Traditionally, we might have classified such an AI system as a form of capital (a software tool) or as an enhancement to human labor (a more productive worker). However, digital labor exhibits qualities that set it apart from both traditional capital and human labor. Digital labor behaves in part like a new type of productive asset and partly like a new type of labor. This section outlines the characteristics of digital labor, contrasting them with human labor and capital. 

\subsection{Intangible \& Non-Physical}
AI is not a machine that you can touch. It exists as code, data, and mathematical weights, an intangible asset without physical form. A state-of-the-art AI model might be composed of hundreds of billions of parameters, hosted across global cloud infrastructure, and continuously refined through model updates and user interaction. Unlike traditional physical assets, their outputs are insights, such as decisions, forecasts, language, and designs, rather than tangible products. This intangibility creates a challenge for traditional accounting and national accounts, which tend to undervalue or omit intangible assets. (We noted above how internal AI software often is not recognized as capital.) However, despite being invisible in accounts, digital labor can be enormously valuable and scalable.

\subsection{Unprecedented Scalability (Non-Rivalry)}
Traditional human labor and capital have natural limits to scale. A human worker can only work so many hours or serve so many customers in a day; a machine can only handle one task at a time in one location. Digital labor, on the contrary, is highly scalable: it can be replicated and deployed at near-zero marginal cost across unlimited processes. Once you develop an AI model, you can deploy it to 1 user or 1 million users with minimal additional cost or degradation. An AI customer service agent can handle 1000 queries as easily as 1 query, which is impossible for a single human agent. In economic terms, digital labor is nonrivalrous: one unit of digital labor (like a trained AI model) can be used simultaneously by many people or for many tasks without diminishing its effectiveness \cite{romer1990techchange}. However, it remains excludable: Although the marginal cost of additional use is close to zero, practical constraints such as licensing, subscription fees, or access controls can restrict who can utilize the AI model. This combination of non-rivalry and practical excludability gives digital labor its distinctive scalability profile, enabling powerful network effects and potential explosive economic growth. 

\subsection{Autonomous Learning \& Self-Improvement}
Perhaps the most novel aspect of AI as a factor is its ability to improve itself through use. Human labor improves over time through training, feedback, and trial-and-error, but learning is slow, bounded by biology, and expensive to scale. Physical capital, on the contrary, degrades with use. As Robert Hall observed, 'a lathe never sharpens itself' \cite{hall2001capital}. AI, in contrast, can. With the right data, usage context, and human guidance, modern AI systems can refine their models, adapt strategies, and deliver increasingly accurate or insightful output. In effect, the marginal productivity of digital labor can increase endogenously with utilization. For example, an AI-powered code assistant might learn from every piece of code it helps generate; over time, it provides faster and more accurate suggestions as it 'learns' from user feedback. This self-improvement means that digital labor potentially offers increasing returns in certain contexts, the opposite of the diminishing returns we expect from adding more traditional labor or capital. It also means that organizations that deploy digital labor can benefit from network effects. The more they use and refine their AI, the more productive it becomes, creating a virtuous cycle of improvement. However, this same dynamism is also a liability. 

\subsection{Rapid Depreciation and Obsolescence}
Unlike traditional capital, whose depreciation follows a relatively stable and predictable path, AI can deteriorate rapidly - sometimes invisibly - due to two mechanisms. First, the informational basis of digital labor can quickly degrade as data becomes outdated, training signals diminish, or models retrain on their synthetic outputs, leading to performance deterioration analogous to mechanical wear \cite{shumailov2024collapse}. Second, digital labor is highly susceptible to rapid obsolescence. Algorithmic breakthroughs (such as model distillation, advanced reinforcement learning, or new prompt engineering methods) can instantly devalue existing models, making recent systems economically obsolete almost overnight. Thus, treating digital labor merely as another type of capital fundamentally misunderstands these dynamics. The same system that compounds value through learning can just as easily slide into irrelevance without active stewardship. This volatility is precisely what separates digital labor from both capital and traditional labor and what makes it such a distinct and strategically complex production factor.

\subsection{Elastic and Nuanced Substitutability with Human Labor}
AI’s relationship to human labor is highly elastic and deeply context dependent. In some tasks, AI fully substitutes for people acting as digital labor. In others, it complements and amplifies human work. One system might automate production scheduling entirely, while another supports a designer with ideation, boosting productivity without replacing the worker. Economists describe this variation as the elasticity of substitution, which differs widely by task. In structured, rules-based environments, digital labor may act as a near perfect substitute, often faster, cheaper, and more reliable. For example, recent studies show that large language models include the correct diagnosis in their differential in 88\% of cases, comparable to clinician performance \cite{levine2023gpt3} and achieve 80\% precision in visual skin condition identification when combining text and images \cite{zhou2024skingpt}. In creative, interpersonal, or judgment-intensive domains, digital labor is more likely to serve as a collaborator than a replacement. This behavior diverges from traditional capital and labor. Although machines have long replaced repetitive physical labor, they have also served as powerful complements in areas such as communication, computation, and transportation, expanding the scope of what humans can achieve. Digital labor extends this logic further; it scales instantly, performs cognitive work, and operates across tasks simultaneously. One system might replace the output of many workers, or a single worker may coordinate multiple AI agents. This leads to a non-linear substitution dynamic; AI adoption is not a one-to-one labor replacement. It can massively multiply output in some areas, leave others untouched, and create entirely new kinds of work. Offloading tasks can make the remaining work more productive, especially when that work involves scarcer forms of expertise. However, there is no guarantee that the remaining tasks will increase in value. If they’re commodified or poorly compensated, wages and employment can fall even as productivity rises. In this sense, digital labor acts not just as a tool, but as a task reallocation engine, reshaping how work is divided, what skills are scarce, and how expertise is rewarded. Digital labor has its own cost structure and strategic trade-offs. However, unlike traditional input, its substitutability with humans is more elastic, more asymmetric, and far more dynamic.

\subsection{Why Digital Labor Deserves Independent Recognition}
Because of these properties, neither classic capital nor classic labor fully encapsulates what digital labor represents. It is created through investment (like capital), but it behaves more like accumulated knowledge that can replicate without physical limits (like software or data). It performs work (like human labor), but it does not tire, can self-improve, and does not receive wages - its “compensation” is the return on investment for its owner. This has profound implications for how we model growth and how companies organize. 

Economic history saw a similar conceptual shift with human capital. In the mid-20th century, economists like Gary Becker and Theodore Schultz argued that education and skills should be treated as capital because they clearly affected output. Initially resisted, the idea was eventually adopted and improved our understanding of growth. By 1992, the augmented Solow model explicitly included human capital alongside physical capital and labor, yielding much clearer insights into national growth differences \cite{mankiw1992growth}. 

By analogy, it may be time to explicitly include digital labor as a factor. Some might argue that digital labor is just another technology investment and subsequently should be left in capital or TFP. However, modeling it separately allows us to understand and manage its unique role in productivity. It makes visible how much of our growth comes from this new factor and how it interacts with other inputs. 

From a strategic business perspective, treating digital labor as a distinct factor means managing it differently. It is not just part of your IT capital budget or a simple efficiency tweak to the workforce. It’s a resource to accumulate (through data, models, training), nurture (through updates, learning), and allocate optimally, just like people and money. It also means tracking its performance on its own terms (how is our digital labor contributing to output?) rather than only measuring downstream effects.

To clarify the distinctiveness of digital labor, Table 1 summarizes key economic properties, explicitly contrasting digital labor with traditional capital and human labor.

\begin{table}[htbp]
\centering
\caption{Comparative Economic Properties of Capital, Human Labor, and Digital Labor}
\begin{tabular}{ |p{3.5cm}||p{3.5cm}|p{3.5cm}|p{3.5cm}| }
 \hline
 Property & Capital & Human Labor & Digital Labor \\
 \hline
Physical vs. Intangible & Physical (machines, hardware); intangible (software) & Physical (people, bodies) & Intangible (models, code, embeddings); outputs cognitive tasks, insights, and decisions \\
 \hline
Scalability / Rivalry & Mixed: physical capital is rivalrous; intangible capital is non-rivalrous but excludable & Rivalrous; limited scalability, constrained by human availability & Non-rivalrous (near-zero marginal cost for additional users) but practically excludable via licensing or access controls \\
 \hline
Improves Over Time? & No; depreciates with use & Yes, but slowly and expensively via training & Yes; self-improving through use, data, and feedback; potential for compounding productivity gains \\
 \hline
Depreciation Profile & Predictable, physical wear & Aging, skill atrophy, illness & Volatile; can decay from data drift or become obsolete rapidly due to training/inference innovation \\
 \hline
Task Substitution & Replaces manual, repetitive tasks & Performs cognitive, emotional, and physical labor & Can substitute or complement depending on task; nonlinear, elastic substitution effects \\
 \hline
Marginal Cost of Replication & High; each new unit costs money & Infinite cost (you can’t clone people) & Near-zero; duplication of models is nearly costless \\
 \hline
Coordination \& Management Needs & Centralized, mechanical operation & High—requires supervision, motivation, development & Medium—requires monitoring, prompt engineering, data curation, updates; acts as a task allocator and autonomous agent \\
 \hline
How Measured in Accounts & Visible: depreciated asset & Partially visible: wages, benefits & Often invisible: underrepresented in financial statements; underestimated in standard productivity measures \\
 \hline
\end{tabular}
\end{table}

\section{Integrating Intelligence into Growth Models: Implications for Business Strategy}
What happens when we take this factor of digital labor and plug it into the economic frameworks that leaders use (explicitly or implicitly) to guide strategy? Classic growth theory gives us two useful lenses: Robert Solow’s model \cite{solow1957technical}, which focuses on capital accumulation and an unexplained tech residual (TFP) driving output; and Paul Romer’s \cite{romer1990techchange} model, which brings innovation and knowledge accumulation into the equation as deliberate growth drivers. By integrating digital labor into these models, we can gain powerful insights for business strategy. 

At a high level, the addition of digital labor as a new productivity factor means that company growth will increasingly depend on how effectively it develops and deploys digital labor alongside human labor and traditional capital. It also reveals two distinct ways that digital labor impacts growth.

\begin{enumerate}
    \item \textit{Boosting productivity in current operations (Solow-type effect)}: Here, digital labor acts as a distinct, scalable input that enhances existing labor and capital efficiency. It can increase output immediately by automating tasks, reducing waste, optimizing decisions, and more. In Solow’s terms, this could increase both the production level and potentially the growth rate during the transition as more digital labor accumulates. But unlike a one-time capital addition, digital labor can also sustain ongoing growth through continuous improvement (learning-by-doing) and by enabling reinvestment into more digital labor (creating a feedback loop) \cite{arrow1962learning}. This suggests strategies around adopting digital labor in production and reinvesting gains.
    \item \textit{Accelerating innovation and knowledge creation (Romer-type effect)}: Here, digital labor explicitly functions as a direct input into the innovation process, performing cognitive tasks traditionally done by researchers. By generating new ideas, designs, or insights faster and on larger scales than humans alone, digital labor increases the efficiency and effectiveness of R\&D processes. In Romer’s framework, more effective 'research labor' (which digital labor can provide) leads to a higher rate of knowledge accumulation, boosting the long-term growth rate of the firm or economy. Strategically, this highlights the importance of explicitly deploying digital labor in product development, R\&D, and creative innovation tasks to maintain competitive advantage and continuous growth.
\end{enumerate}

\subsection{Re-imagining Resource Allocation (Human Labor + Capital + Digital Labor)}
An immediate strategic implication is that leaders must optimize the mix of \textit{three} inputs –capital and human labor and digital labor (AI) - not just two. This means asking new questions: In which tasks or processes should we deploy AI, and where should we rely on human expertise? and How much budget should shift from traditional capital expenditures (such as equipment) or labor costs into developing and maintaining AI systems?

\begin{figure}[htbp]
\centering
\includegraphics[width=\textwidth]{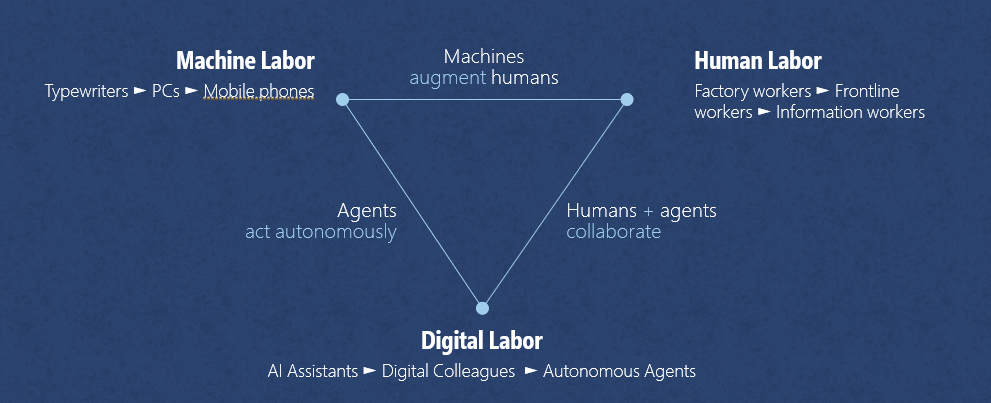}
\caption{Illustrates the strategic reallocation framework in an AI-driven economy, emphasizing the relationships and complementarities among machine labor, human labor, and digital labor (AI). It highlights the dynamic interactions and varying degrees of autonomy across these three inputs, reinforcing the necessity for organizations to thoughtfully map tasks and processes to the most suitable combination of labor types.}
\end{figure}

Effective allocation in an AI-driven economy involves mapping the cognitive work in an organization to the most appropriate type of labor or automation.
\begin{itemize}
    \item Tasks requiring creativity, complex judgment, moral reasoning, or empathy remain human-centered, possibly augmented -but not replaced - by digital labor.
    \item Routine, data-intensive, or repetitive cognitive tasks are prime candidates for digital labor automation.
    \item Predictable physical tasks remain suitable for traditional automation (robots or machinery).
    \item Emergent roles around managing or enhancing digital labor itself, such as data curation, model oversight, and prompt engineering, represent new forms of human-AI complementarity.
\end{itemize}
As technologies advance and societal expectations evolve, this allocation will continue to shift, demanding ongoing strategic reevaluation \cite{susskind2025futurework}.

\section{Rethinking Productivity and Performance Metrics}
Considering digital labor as a third production factor means that the measurement of internal productivity must evolve. Business leaders should establish metrics that specifically capture the contribution of digital labor. This might involve tracking AI system output (e.g. queries resolved, accuracy, cycle time reduction) as separate dashboard items, not just attributing results to traditional departments. For example, if digital labor acting as a scheduler increases manufacturing throughput by 15\%, that gain should be credited to the AI system, measuring the marginal product of intelligence. Comparing this to metrics such as output per human worker provides a more complete picture. 

Some companies are introducing AI productivity metrics that track the value generated (revenue increase, cost savings, or quality improvements). This justifies the investment in digital labor and identifies underperforming initiatives. Furthermore, firm-level TFP (the residual in productivity analysis) can be better understood by identifying the role of AI.

Another crucial metric is the learning/improvement rate of AI systems. Since digital labor can improve over time, companies might track metrics such as model accuracy or task completion time over successive iterations. Plateauing performance might signal the need for updates and improvement signals the compounding value. This parallels tracking employee development and productivity improvements after training, but applies to AI.

Integrating digital labor means integrating it into measurement systems. Leaders expand KPIs to include digital labor contribution and indicators of intangible assets. This makes the impact of digital labor visible in reports, preventing the 'productivity paradox' (investing in AI without seeing results in numbers) \cite{brynjolfsson2017paradox}.

\section{Organizational Design for Human–AI Innovation}

Integrating digital labor into the enterprise is not simply a matter of deploying a new software tool. Digital labor challenges the very architecture of how organizations function, how they generate knowledge, distribute expertise, and make decisions. Treating digital labor as a standard IT investment risk under utilizing its capabilities and misaligned incentives. Instead, unlocking its full value requires deliberate organizational design: restructuring teams, workflows, and roles to enable effective collaboration between humans and AI systems.

Crucially, this is not just about upgrading your technology stack; it is about investing in human labor alongside it. Humans are what moves organizations out of equilibrium. They decide what is worth doing, break routines, find purpose, and invent new problems to solve, things that digital labor cannot do. The most effective digital labor strategies pair automation with augmentation, ensuring that people continue to provide creative direction, strategic oversight, and human judgment. That requires intentional investment in the skills, processes, and governance structures that make human–digital labor collaboration not just possible, but productive and trustworthy.

This includes:
\begin{itemize}
    \item \textit{Focus on data quality and knowledge management:} High-quality data and institutional knowledge are critical to the effectiveness of digital labor. Build robust data pipelines and knowledge repositories accessible to both humans and digital labor. This cultivates the knowledge stock that fuels compounding returns.
    \item \textit{Create new roles/processes: }Adapt structure with roles like AI trainers, data curators, and processes for validation and collaboration
    \item \textit{Mind the talent factor:} As digital labor handles routine cognitive tasks, high-level creative and strategic human skills become more valuable. The return on exceptional talent increases as their insights can be magnified by AI. Invest in attracting, retaining and enabling top talent with AI.
    \item \textit{Govern digital labor implementation:} Because AI performance is tied to the quality data and processes generated and overseen by these talented individuals, establish strong oversight. This includes practices such as regular audits, incorporating the diverse points of view of your skilled teams, and ensuring that digital labor objectives are clearly aligned with business values to promote a successful and trustworthy implementation.
\end{itemize}

In summary, designing for an AI world means integrating digital labor while improving human creativity, data quality, and learning. Successful companies will likely create a 'flywheel' of human AI that compounds: humans build better AI \texttt{-->} AI augments humans \texttt{-->} humans create new knowledge \texttt{-->} that trains better AI \texttt{-->} driving competitive advantage and sustained growth. 

\section{A New Equation for Strategic Success}
Artificial intelligence may be reshaping the very foundations of the productivity equation. By introducing a scalable, self-improving, and increasingly autonomous form of labor, AI raises new questions about how value is created and by whom. If digital labor doesn’t just enhance existing inputs but alters the structure of production itself, should we begin to think of it not as a tool, but as a distinct factor of production? And if so, what would it mean to develop, measure, and manage it alongside capital and human labor? 

Updating traditional metrics and growth models to account for digital labor reveals both risks and opportunities. The risk is that by clinging to outdated frameworks, relying on incomplete metrics, overlooking AI’s contribution, or treating it merely as capital, organizations will misjudge performance and underinvest in the true drivers of competitiveness. The opportunity lies with those who embrace this emerging input. Firms that integrate AI alongside human labor and physical capital can unlock immediate efficiencies, sustained innovation-led growth, or both, ultimately leading to higher profits and a more agile and future-ready organization. 

\section*{Acknowledgments}
This paper benefited from the insights, feedback, and collaboration of many colleagues. We are especially grateful to Sonia Jaffe, Michael Schwarz, Donald Ngwe, Jaime Teevan, and Brent Hecht for their partnership in refining these ideas, as well as to the broader Microsoft Research community for thoughtful critique and support. We also thank the researchers and practitioners whose work continues to shape our understanding of digital labor and the future of productivity. 

\bibliographystyle{apalike}
\bibliography{references}

\end{document}